# Micro-transfer printing high-efficiency GaAs photovoltaic cells onto silicon for wireless power applications

Ian Mathews[1],*, David Quinn[1],*, John Justice[1], Agnieszka Gocalinska[1], Emanuele Pelucchi[1], Ruggero Loi[1], James O'Callaghan[1], and Brian Corbett[1],^

[1]Tyndall National Institute, Lee Maltings, University College Cork, Ireland

\* Equal contribution
^ Corresponding author (brian.corbett@tyndall.ie)

**Abstract**

Here we report the development of high-efficiency microscale GaAs laser power converters, and their successful transfer printing onto silicon substrates, presenting a unique, high power, low-cost and integrated power supply solution for implantable electronics, autonomous systems and internet of things applications. We present 300 μm diameter single-junction GaAs laser power converters and successfully demonstrate the transfer printing of these devices to silicon using a PDMS stamp, achieving optical power conversion efficiencies of 48% and 49% under 35 and 71 W/cm$^2$ 808 nm laser illumination respectively. The transferred devices are coated with ITO to increase current spreading and are shown to be capable of handling very high short-circuit current densities up to 70 A/cm$^2$ under 141 W/cm$^2$ illumination intensity (~1400 Suns), while their open circuit voltage reaches 1235 mV, exceeding the values of pre-transfer devices indicating the presence of photon-recycling. These optical power sources could deliver Watts of power to sensors and systems in locations where wired power is not an option, while using a massively parallel, scalable, and low-cost fabrication method for the integration of dissimilar materials and devices.

**Introduction**

Micro-transfer printing (μTP) provides a means of monolithically integrating dissimilar materials and components on non-native substrates that is scalable, massively parallel and potentially low-cost [1]. The transfer print process, by combining diverse optical, electronic and other functional materials, opens up an enormous range of possibilities for new devices with embedded functionality leading to more compact chips and systems for a variety of applications, such as telecommunications, smart sensing, biomedical sensing and data storage. For III-V materials and devices, anywhere from 1 to 10,000's of components can be removed from their growth substrates and printed systematically and simultaneously on to a chosen substrate, e.g. silicon, with automated micron-scale alignment [2]. Micro-transfer printing presents a compelling alternative to address the challenge of integrating non-compatible components in large volumes at the semiconductor wafer level, providing many advantages over current photovoltaic integration processes such as wire-bonding, wafer-bonding and direct growth [3], [4].

There are a wide range of embedded systems, with examples provided in Figure 1, that benefit from the integration of dissimilar materials and devices but operate in difficult environments where providing electrical power through wiring or batteries is a challenge such as implantable biomedical sensors and devices [5], 5G networks [6], where voltage isolation is required such as for lightning safe wind turbine blade monitoring [7], underwater robotics [8] and many military applications including airborne or submerged vehicle charging and satellite or sensor charging in hazardous locations [9]. Energy harvesting solutions such as implantable piezo-electrics [10], photovoltaics for wireless sensors [11], [12] or subdermal systems [13], [14] and radio-frequency harvesting [15] have all been demonstrated as power sources for autonomous systems and sensors. Optical power transfer is an alternative technology that rather than relying on energy harvested from the immediate environment, uses a photonic source such as a laser to send optical power either through free space or over an optical fiber to remotely powered autonomous systems. Micro-transfer printing laser power converters offers many advantages over environmental energy harvesting systems including the ability to directly integrate the device with silicon electronics, miniaturization of the power supply for microscale systems [16], and the utilization of only the few micron thick essential layers of the photovoltaic device and the re-use of the growth substrate to reduce cost, as was previously demonstrated for GaAs photovoltaics through epitaxial lift-off [17]. In addition, the power supplied by environmental energy harvesters is typically restricted to the micro-watt range, limiting the functionality of the autonomous system whereas mm-scale laser power converters are capable of supplying Watts of power. Transfer printing micro-PV cells for high intensity optical power conversion has been successfully demonstrated for multi-junction solar cells under concentrated sunlight where the final microcells had a conversion efficiency of 43.9% at concentrations exceeding 1,000 Suns (100 W/cm$_2$) [18].

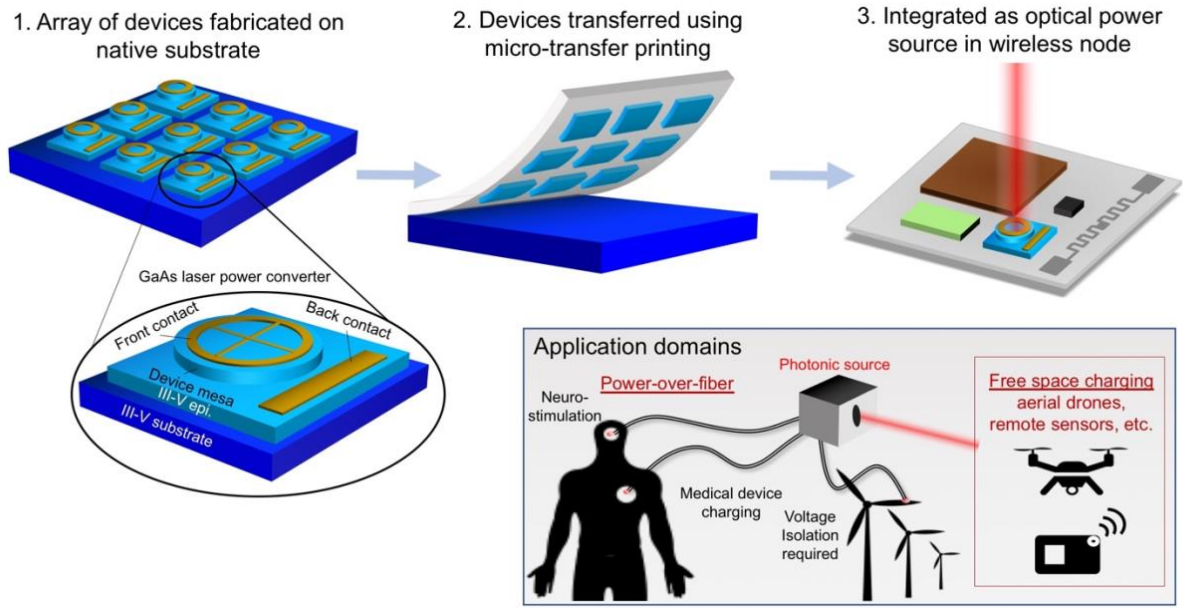

Figure 1: A schematic summary of the fabrication and micro transfer printing of on-chip III-V optical power sources and the potential applications of the technology including power-over-fiber and free space optical charging.

GaAs photovoltaic cells have been widely studied as laser power converters and currently hold the record conversion efficiency of 66.3%, under 150 W/cm$_2$ illumination at a wavelength of 837 nm, using a monolithic lattice-matched device with 20 GaAs junctions vertically stacked in series [19]. Fafard *et al.* [19] and Masson *et al.* [20] have developed these vertical epitaxial heterostructure architectures (VESHA) consisting of many thin n-on-p GaAs photovoltaic junctions engineered to optimize absorption and current matching between the junctions. This approach requires precise epitaxial growth as the junctions are grown in a single crystal with, e.g., the utilization of InGaP lattice matched window and back surface field layers. Khvostikov *et al.* [21] hold the current record conversion efficiency for a single junction GaAs based PV converter. They describe a p-on-n structure with an AlGaAs window and back surface field layers where their final devices achieved 60% at a photocurrent density of 5.9 A/cm$_2$ under 809 nm monochromatic illumination with a fill factor of 83.9% and a $V_{oc}$ of 1.19 V.

In the rest of this paper, we present the first transfer printing of GaAs photovoltaic microcells to silicon for laser power conversion applications. We begin by introducing our micro-transfer printing process and describe the fabrication of high-efficiency single-junction GaAs laser power converters. Transfer printing of these devices is successfully demonstrated using PDMS stamps after the release of the device coupons using photoresist tethers and an undercut etch process. We then compare the performance of these devices before and after transfer-printing onto silicon substrates, with the devices on silicon maintaining higher power conversions at high incident power, while their open circuit voltage exceeds the values of pre-transfer devices indicating the presence of photon-recycling [22]. Next, using electroluminescence measurements we describe how coating the transferred devices with an indium-tin oxide lateral conduction layer increases current spreading and enables them to be capable of handling very high short-circuit current densities up to 70 A/cm$_2$. In the final section we describe our optimized optoelectronic devices on silicon where the use of an ITO spreading layer requires the optical properties of the transferred device stack to be further improved by the addition of an optimized SiN anti-reflection coating leading to 5% reflectance at 800 nm, with a peak EQE of 89% at 808 nm, the operating wavelength. The devices open circuit voltage reaches 1235 mV and absolute power conversion efficiencies of 49%, 48% and 44% are measured under 35, 71 and 141 W/cm$_2$ 808 nm laser illumination respectively, post transfer printing to silicon.

**Results and discussion**

Previously, to integrate many materials and components from various substrates onto a single platform, serial pick and place assembly operations were considered the primary option. There are many potential flaws with these systems. Standard pick and place assembly struggles with small (<100 μm edge) and thin-film devices. Modern pick and place assemblies can operate at high rates, but with the penalty of a reduction in placement accuracy of components making it a less viable option for the integration of state-of-the-art devices as they reduce in size. Micro-transfer printing provides a powerful means to heterogeneously integrate many components onto a single target wafer.

For III-V materials, an elastomer stamp is used to pick the few microns thick essential layers of a device and print them to a target substrate. An etch release layer is incorporated into the epi-structure of the device between the substrate and the active epitaxial layers. The release layer is designed to be selectively etched with a wet chemical etchant while lithographically defined photoresist tethers anchor the active material to the substrate and keep the device suspended above the substrate. A computer-controlled elastomer stamp is then used to pick the suspended device from the host substrate causing the tethers to fracture at specifically engineered points allowing the devices to be printed to a target substrate of choice.

The device fabrication is summarized here and in Figure 2. In this work, a GaAs based PV converter was grown on a GaAs wafer by metal organic vapor phase epitaxy (MOVPE) in an Aixtron 200 horizontal MOVPE reactor [23] with a p-on-n epi-structure consisting of a thin GaAs emitter and thick GaAs base with AlGaAs top and rear current spreading layers – the complete layer structure is provided in Table 2 in the Experimental Methods Section. The first step is to form the device mesa and front and back electrical contacts as shown in Figure 3 (a). A 500 nm thick AlInP etch release layer is provided as the first layer in the epi-structure to allow the undercut of the devices with a HCl:H$_2$O 1:1 etch. To enable the systematic release and transfer of the devices, a photoresist tethering system was used that allows the etch of the release layer while the devices are suspended and held in their original location once undercut. This allows registered pick and the print of the devices on to the target silicon substrate with a PDMS elastomer stamp. In Figure 3 (b), the remaining 'broken' tethers on the GaAs substrate are clearly visible post-transfer printing.

The increased resistance due to the micro-transfer print process must be addressed if the potential benefits of transfer printing for PV devices are to be utilized. The resistive non-ohmic path at the rear of the devices prevents the current from utilizing the substrate for lateral conduction. In order to reduce electrical resistance in this direction, gold germanium nickel (AuGeNi) was evaporated on to the rear of the devices during the transfer print process. Devices were picked using a PDMS stamp (Figure 3 (c)), the entire PDMS stamp was then placed into a Temescal FC2000 Electron Beam evaporator where a thin-film AuGeNi layer was then evaporated onto the rear of the picked devices. AuGeNi is known to provide an ohmic contact to n-type GaAs while it allows rapid heat dissipation from the device due to gold's large thermal conductivity (200 W/mK for a 100 nm thin film on Si). The devices were then printed on to gold coated silicon with the aid of a hot plate. The photoresist was stripped and the printed devices were annealed in a furnace at 350 °C for 15 minutes. To further remedy the current crowding effects, a 100 nm Indium Tin Oxide (ITO) transparent conduction layer was evaporated on the front surface of some devices before transfer printing. ITO was chosen at it has low resistivity while also having high transmission in the infrared wavelength regime (i.e. 808 nm). The undercut, pick, and printing of the GaAs PV cells was successfully applied to ~500 μm x 500 μm coupons with an array of transferred devices with multiple front-contacting schemes shown in Figure 3 (d).

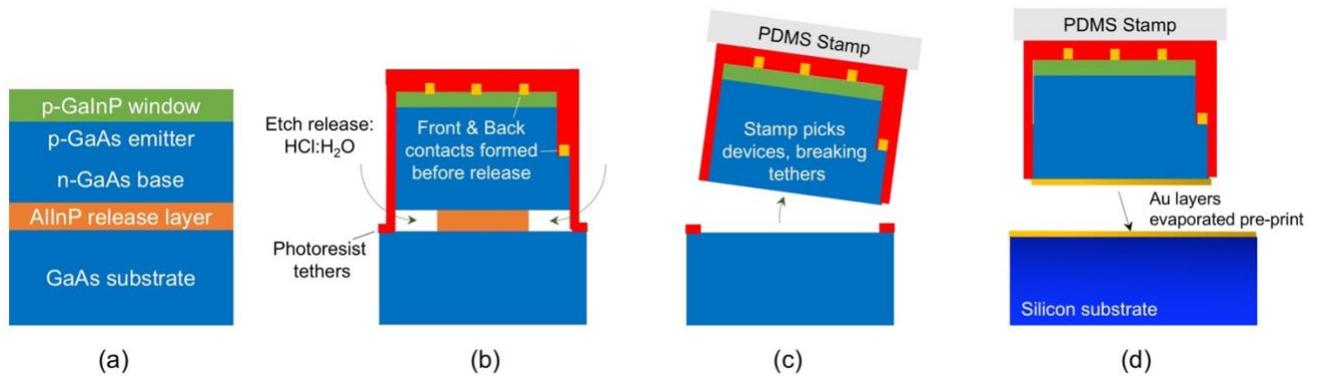

Figure 2: Schematic summary of the device fabrication and micro-transfer printing process including; (a) the initial III-V photovoltaic structure grown by metal-organic vapor phase expitaxy on a GaAs substrate, (b) the device mesas are formed and the front and back contacts deposited before a photoresist tether system and etch release used to remove the AlInP release layer, (c) the devices held by the tethers are released from the growth substrate by mechanically picking them with a PDMS stamp and, (d) the devices are printed onto a gold-coated silicon substrate after the device back surface is also coated with a Au layer that aids the printing process and reduces lateral resistance in the devices.

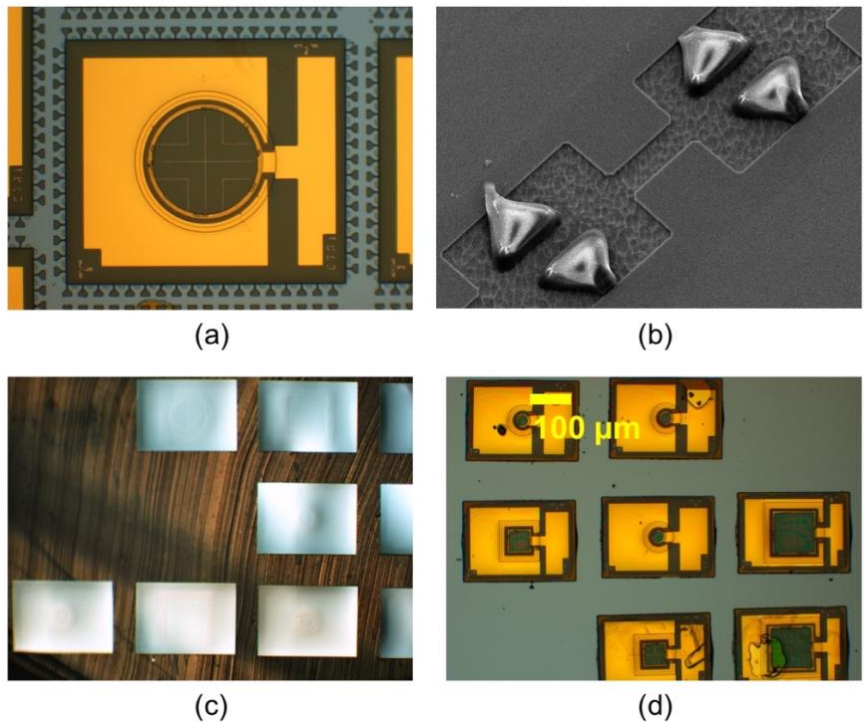

Figure 3: Images of the fabrication and micro-transfer printing process including (a) a fabricated device including front and back contacts pre-transfer printing, (b) scanning electron microscopy image of the 'broken' resist tethers on the GaAs substrate post-device release, (c) the underside of a PDMS stamp with coupons attached post-removal from the native GaAs substrate and, (d) a number of laser power converters micro-transfer printed onto silicon.

In this section we compare the illuminated current-voltage characterization of laser power converters on-chip (pre-transfer print), on-chip with an ITO current spreading layer, and post-transfer print to silicon. Given the difference in the optical and electrical structures of these devices we compare the normalized fill factor and efficiency versus light intensity in order to analyze trends rather than compare results directly – for each device the measurements are normalized to the measurements taken at the lowest incident power intensity of 14 W/cm$_2$.

A fiber-coupled 808 nm multimode laser was used to illuminate the PV devices where the exit aperture of the fiber was 200 μm in diameter with a numerical aperture of 0.2. On-chip performance with and without an ITO spreading layer are compared to devices with an ITO spreading layer and have been transfer printed to gold-coated silicon as outlined in Figure 4. Figure 4 (b) presents the fill factor of each device versus light intensity and shows how standard on-chip devices suffer from a significant drop as the limited current spreading in the devices leads to higher series resistance at high currents and a steep decline in *FF*. To address this current spreading issue, a 100 nm ITO spreading layer was coated on the top of the devices as described above. The *FF* and *PCE*, shown in Figure 4 (b) & (c), of these on-chip devices now sustain higher performance to larger incident power levels maintaining values between 70% – 80% of the 14 W/cm$_2$ value in the 30 – 150 W/cm$_2$ range, and *FFs* above 90% of the 14 W/cm$_2$ value. The *Jsc* of the devices (not-shown), however, is lower than the on-chip devices without ITO owing to the increased reflectance and parasitic absorbance caused by the addition of the thin-film layer. Finally, the performance of the micro-transfer printed devices was measured and was shown to maintain the *PCE* across the range of light intensities. The *FF* of the devices drops more than the ITO-coated on-chip devices, owing to an increase in resistance, but the boost in $V_{oc}$ observed (Figure 4(d)) leads to higher *PCEs* across the light intensity range. The transferred devices include the front surface ITO and a rear surface AuNiGe spreading layers. The AuNiGe spreading layer also acts as a partial rear reflector and increases photon absorption in the devices. This increased carrier concentration leads to a subsequent increase in open-circuit voltage due to photon-recycling, and results in an exceptionally high open-circuit voltage of 1230 mV measured under 141 W/cm$_2$ illumination, 25 mV and 40 mV above the values measured for the on-chip devices with and without ITO respectively, and greater than the highest performing GaAs laser power converters in the literature as described in Table 1.

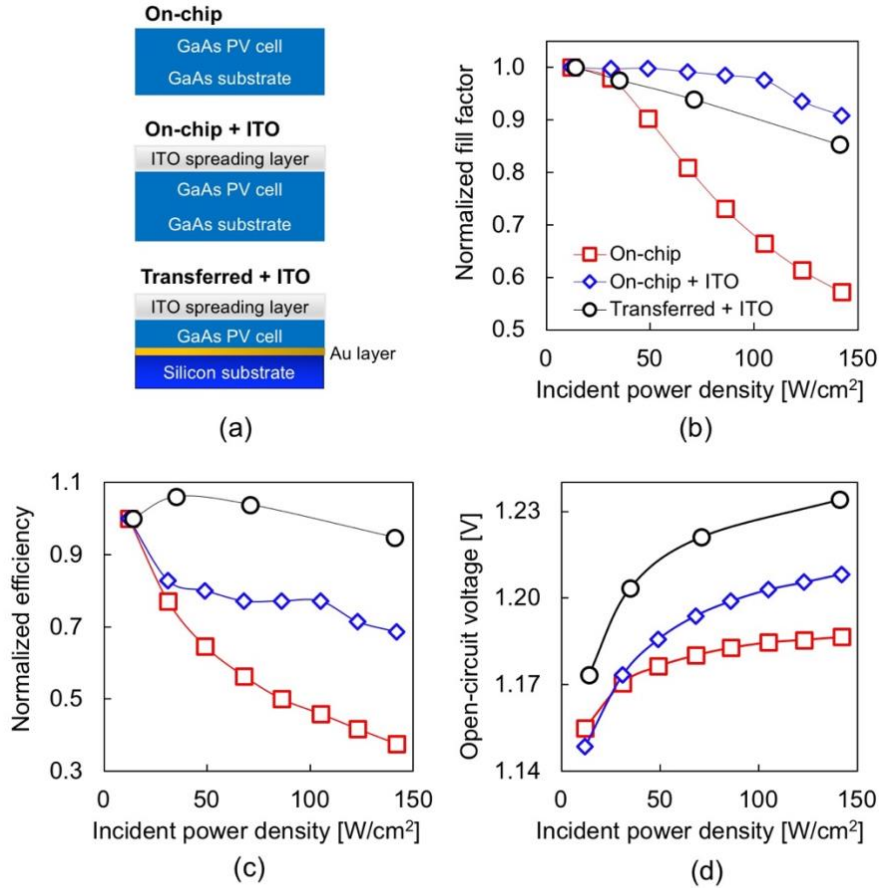

Figure 4: (a) Schematic outline of the layer structures of the three device types compared, (b) normalized fill factor, (c) normalized PCE, and (d) open-circuit voltage, all versus incident power density of the optical source.

The results of the illuminated current-voltage measurements showed the benefits of the current spreading layers under high incident power densities. To further analyze current crowding effects in the devices and the impact of the front surface ITO spreading layer only, electroluminescence emission profile measurements were taken on devices with and without this layer as in Figure 5. Figures 5 (b) & (c) present the EL emission profiles of 300 μm diameter on-chip devices with and without the ITO current spreading layer respectively. Initially, under a low current density injection of 10 mA (14 A/cm$^2$) the emission profile of both devices showed homogeneous light emission from the mesa (not shown). As the injection current density was increased to 150 mA, the emission becomes less homogeneous for both devices but this current crowding effect is much more apparent in the device without the ITO current spreading layer where the current is crowding towards the 'left' of the devices where the devices are probed. The metal grid lines do not contribute to the current spreading. Figures 5 (b) & (c) also show the emission profile with a horizontal line scan with relative emission intensity along the red line. These line scans further demonstrate the effectiveness of ITO as a current spreading layer.

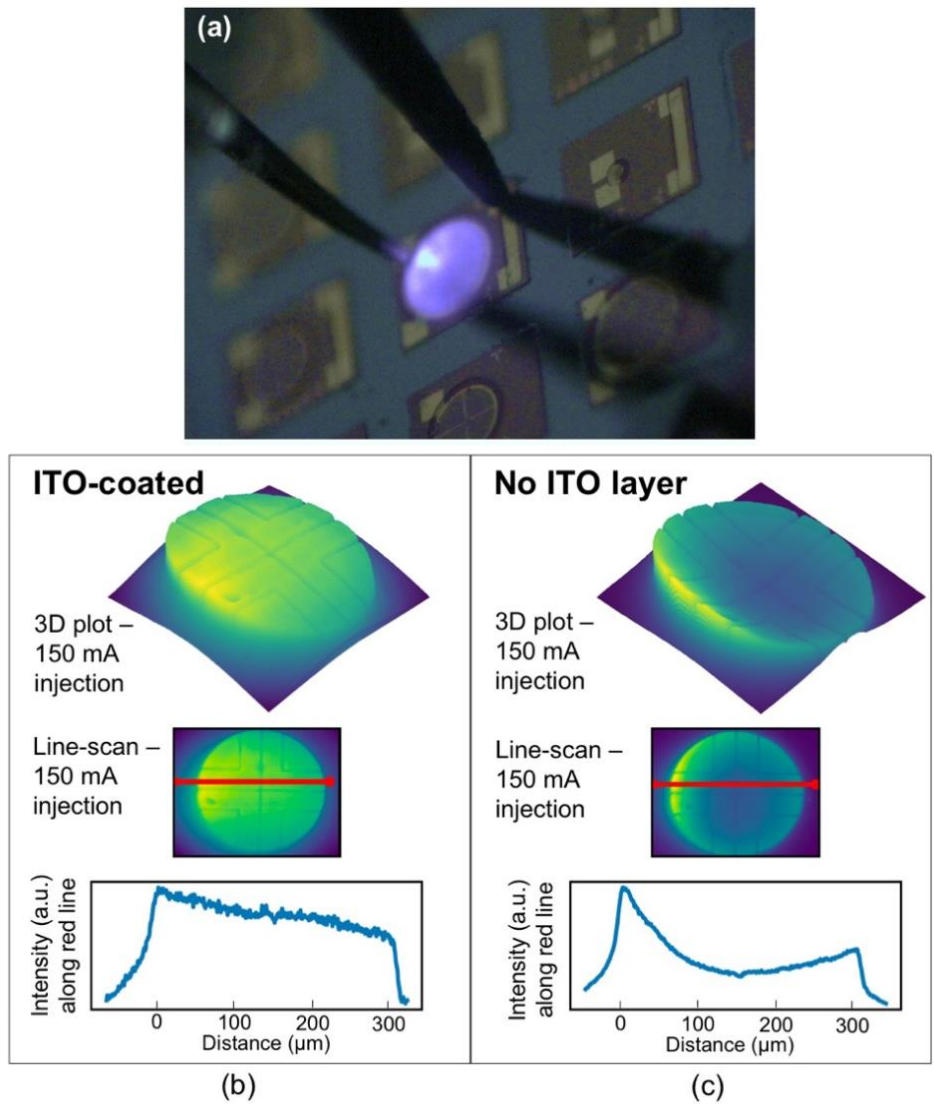

Figure 5: (a) Optical microscope image of a 300 μm diameter device during EL measurements, (b) EL measurements for the on-chip devices for an applied current of 150 mA with an ITO spreading layer, and (c) without an ITO spreading layer.

Following the fabrication and comparison of the on-chip and transfer printed devices, an optimized optical coating scheme was designed to reduce reflectance from the final device that considers the combined interference effects of the ITO and SiN layers on performance and coated on the transfer printed devices. The front surface reflection of the stack was reduced to 5% at 800 nm by depositing a 140 nm of $Si_3N_4$ layer. The external quantum efficiency (EQE) and illuminated current-voltage characteristics of these devices were then measured under varying incident optical power densities and are presented in Figure 6.

The *EQE* of the GaAs epi-structure was measured on large solar cell structures to enable full light collection, and is presented in Figure 6 (c). The devices reach a peak EQE of 91% at 785 nm and 89% at 808 nm, the operating wavelength of the illuminating laser. The EQE differs from that expected for typical solar cells where a much broader response is desired owing to the larger range of wavelengths that are present in the solar spectrum. The measured EQE here has a narrow peak due to the optimization of the anti-reflection coating for 808 nm wavelength, and the absorption of shorter wavelengths in the AlGaAs current spreading layer (The bandgap of $Al_{0.1}Ga_{0.9}As$ is 1.565 eV which corresponds to a wavelength of 792 nm).

The 300 μm diameter single-junction GaAs laser power converters were successfully transfer printed to silicon using a PDMS stamp, achieving optical power conversion efficiencies of 47.8% and 49.4% under 35 and 71 $W/cm^2$ 808 nm laser illumination respectively as shown in Figures 6 (b), (c) & (d). The optimized design of the whole stack for resistance and optical considerations leading to very high efficiency photovoltaic cells on silicon and *FFs* of 85% – 75% across the measurement range. The transferred devices were coated with ITO to increase current spreading and are shown to be capable of handling very high short-circuit current densities up to 70 $A/cm^2$, (Figure 6 (e)) under 141 $W/cm^2$ illumination intensity (~1400 suns equivalent). For devices transferred to silicon, the rear back mirror of AuNiGe resulted in photon recycling, thus increasing carrier confinement and enabled large open circuit voltages exceeding the values of pre-transfer devices, and reaching as high as 1235 mV in the final optimized design.

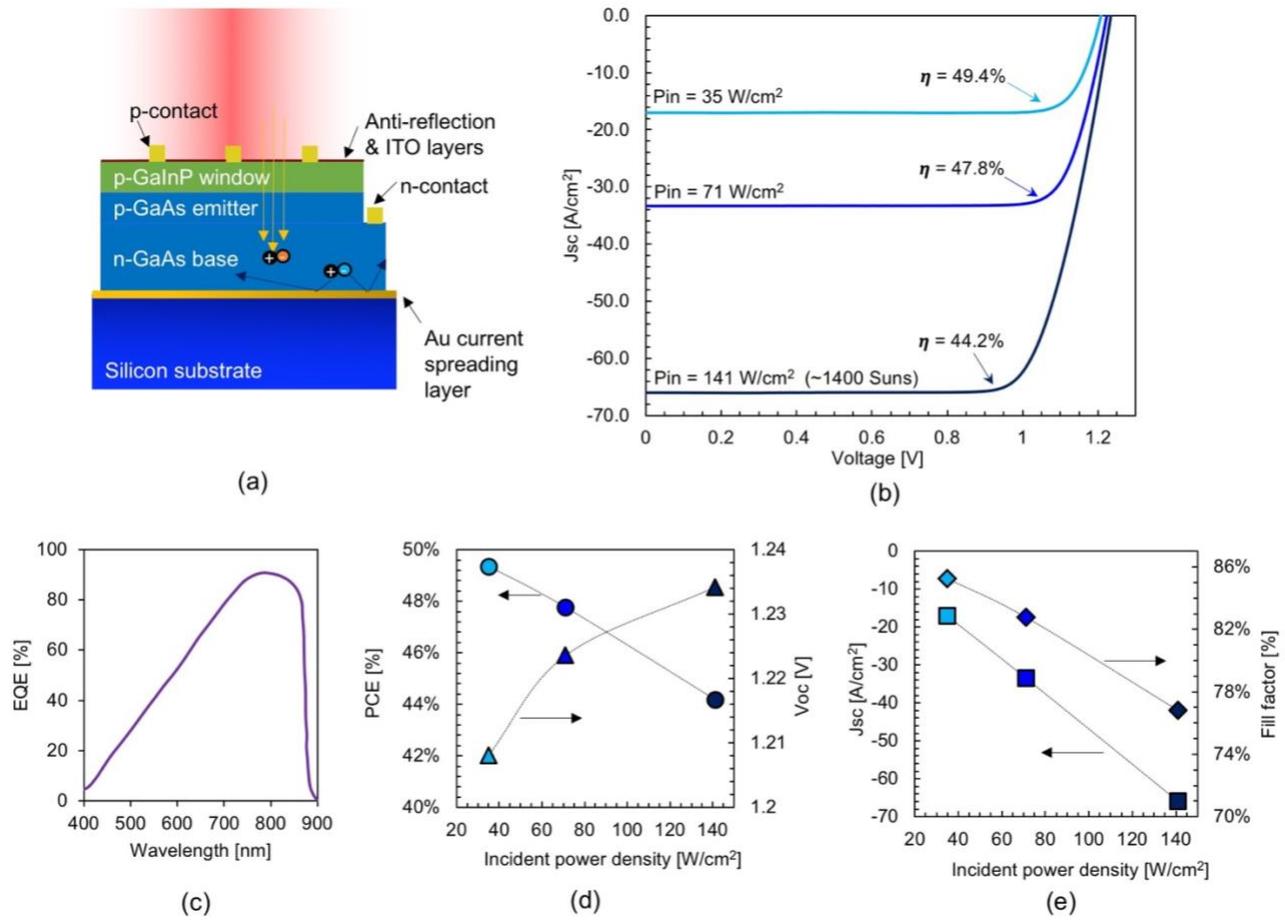

Figure 6: (a) Outline of the final device structure, (b) illuminated current-voltage measurements at three different incident power densities of 35, 71 and 141 W/cm$_2$, (c) the measured photovoltaic conversion efficiency, *PCE,* and open-circuit voltage, *Voc*, and (d) the short-circuit current density, *Jsc* and fill factor, *FF*, of the final devices all versus incident power density of the optical source.

Table 1: A summary of the best performing devices produced in this work and their comparison with the start of the art in GaAs laser power converters.

| Material | $P_{in}$ (W cm$^{-2}$) | PCE (%) | $V_{oc}$/Junction (mV) | Lambda (nm) | FF (%) | Reference |
|---|---|---|---|---|---|---|
| GaAs | 150 | 66.3 | 1.16 | 837 | - | [19] |
| GaAs | 100 | 60 | 1.18 | 809 | 83.9 | [21] |
| GaAs/Si - µTP | 71 | 48 | 1220 | 808 | 83 | *This work* |

## Conclusions

We reported the fabrication of high-efficiency microscale GaAs laser power converters, and their successful transfer printing onto silicon substrates. The 300 μm diameter single-junction GaAs laser power converters were successfully transfer printed to silicon using a PDMS stamp, achieving optical power conversion efficiencies of 48% and 49% under 35 and 71 W/cm$_2$ 808 nm laser illumination respectively. The optimized design of the whole stack for resistance and optical considerations leading to very high efficiency photovoltaic cells on silicon. The transferred devices were coated with ITO to increase current spreading and are shown to be capable of handling very high short-circuit current densities up to 70 A/cm$_2$ under 141 W/cm$_2$ illumination intensity (~1400 suns equivalent). For devices transferred to silicon, the rear back mirror of AuNiGe resulted in photon recycling, thus increasing carrier confinement and enabled large open circuit voltages exceeding the values of pre-transfer devices, and reaching as high as 1235 mV in the final optimized design. Overall, these optical power sources could deliver Watts of power to remote autonomous sensors and systems, without the need for electrical wiring, and using a massively parallel, scalable, and low-cost fabrication method for integrated devices.

## Acknowledgements

The authors acknowledge the sources of funding for this work. I.M. has received funding from the European Union's Horizon 2020 research and innovation program under the Marie Skłodowska-Curie grant agreement No. 746516. The work was also supported by Science Foundation Ireland under Grant Nos. 12/RC/2276, 12/RC/2276-P2, 15/IA/2864 and by the H2020 project TOPHIT.

## References

[1] M. A. Meitl *et al.*, "Transfer printing by kinetic control of adhesion to an elastomeric stamp," *Nat. Mater.*, vol. 5, no. 1, pp. 33–38, Jan. 2006.

[2] J. Justice, C. Bower, M. Meitl, M. B. Mooney, M. A. Gubbins, and B. Corbett, "Wafer-scale integration of group III-V lasers on silicon using transfer printing of epitaxial layers," *Nat. Photonics*, vol. 6, no. 9, pp. 610–614, Sep. 2012.

[3] I. Mathews, D. O'Mahony, K. Thomas, E. Pelucchi, B. Corbett, and A. P. Morrison, "Adhesive bonding for mechanically stacked solar cells," *Prog. Photovolt. Res. Appl.*, vol. 23, no. 9, pp. 1080–1090, Sep. 2015.

[4]   F. Dimroth *et al.*, "Comparison of Direct Growth and Wafer Bonding for the Fabrication of GaInP/GaAs Dual-Junction Solar Cells on Silicon," *IEEE J. Photovolt.*, vol. 4, no. 2, pp. 620–625, Mar. 2014.

[5]   M. Mujeeb-U-Rahman, D. Adalian, C.-F. Chang, and A. Scherer, "Optical power transfer and communication methods for wireless implantable sensing platforms," *J. Biomed. Opt.*, vol. 20, no. 9, p. 095012, Sep. 2015.

[6]   Q. Wu, G. Y. Li, W. Chen, D. W. K. Ng, and R. Schober, "An Overview of Sustainable Green 5G Networks," *IEEE Wirel. Commun.*, vol. 24, no. 4, pp. 72–80, Aug. 2017.

[7]   K. Worms *et al.*, "Reliable and lightning-safe monitoring of wind turbine rotor blades using optically powered sensors," *Wind Energy*, vol. 20, no. 2, pp. 345–360, 2017.

[8]   J. I. de O. Filho, A. Trichili, B. S. Ooi, M.-S. Alouini, and K. N. Salama, "Towards Self-Powered Internet of Underwater Things Devices," *ArXiv190711652 Eess*, Jul. 2019.

[9]   P. Sprangle, B. Hafizi, A. Ting, and R. Fischer, "High-power lasers for directed-energy applications," *Appl. Opt.*, vol. 54, no. 31, pp. F201–F209, Nov. 2015.

[10]  C. Dagdeviren *et al.*, "Conformal piezoelectric energy harvesting and storage from motions of the heart, lung, and diaphragm," *Proc. Natl. Acad. Sci.*, vol. 111, no. 5, pp. 1927–1932, Feb. 2014.

[11]  I. Mathews *et al.*, "Self-Powered Sensors Enabled by Wide-Bandgap Perovskite Indoor Photovoltaic Cells," *Adv. Funct. Mater.*, vol. 29, no. 42, p. 1904072, 2019.

[12]  S. N. R. Kantareddy, I. Mathews, R. Bhattacharyya, I. M. Peters, T. Buonassisi, and S. E. Sarma, "Long Range Battery-Less PV-Powered RFID Tag Sensors," *IEEE Internet Things J.*, vol. 6, no. 4, pp. 6989–6996, Aug. 2019.

[13]  K. Song *et al.*, "Subdermal Flexible Solar Cell Arrays for Powering Medical Electronic Implants," *Adv. Healthc. Mater.*, vol. 5, no. 13, pp. 1572–1580, 2016.

[14]  L. Lu *et al.*, "Biodegradable Monocrystalline Silicon Photovoltaic Microcells as Power Supplies for Transient Biomedical Implants," *Adv. Energy Mater.*, vol. 8, no. 16, p. 1703035, 2018.

[15]  C. M. Boutry *et al.*, "Biodegradable and flexible arterial-pulse sensor for the wireless monitoring of blood flow," *Nat. Biomed. Eng.*, vol. 3, no. 1, p. 47, Jan. 2019.

[16]  E. Moon, I. Lee, D. Blaauw, and J. D. Phillips, "High-efficiency photovoltaic modules on a chip for millimeter-scale energy harvesting," *Prog. Photovolt. Res. Appl.*, vol. 27, no. 6, pp. 540–546, 2019.

[17]  W. Choi *et al.*, "A Repeatable Epitaxial Lift-Off Process from a Single GaAs Substrate for Low-Cost and High-Efficiency III-V Solar Cells," *Adv. Energy Mater.*, vol. 4, no. 16, p. 1400589, 2014.

[18]  X. Sheng *et al.*, "Printing-based assembly of quadruple-junction four-terminal microscale solar cells and their use in high-efficiency modules," *Nat. Mater.*, vol. 13, no. 6, pp. 593–598, Jun. 2014.

[19]  S. Fafard *et al.*, "High-photovoltage GaAs vertical epitaxial monolithic heterostructures with 20 thin p/n junctions and a conversion efficiency of 60%," *Appl. Phys. Lett.*, vol. 109, no. 13, p. 131107, Sep. 2016.

[20]  D. Masson, F. Proulx, and S. Fafard, "Pushing the limits of concentrated photovoltaic solar cell tunnel junctions in novel high-efficiency GaAs phototransducers based on a vertical epitaxial heterostructure architecture," *Prog. Photovolt. Res. Appl.*, vol. 23, no. 12, pp. 1687–1696, 2015.

[21]  V. P. Khvostikov *et al.*, "Photovoltaic laser-power converter based on AlGaAs/GaAs heterostructures," *Semiconductors*, vol. 50, no. 9, pp. 1220–1224, Sep. 2016.

[22]  X. Sheng *et al.*, "Device Architectures for Enhanced Photon Recycling in Thin-Film Multijunction Solar Cells," *Adv. Energy Mater.*, vol. 5, no. 1, p. 1400919, 2015.

[23]  V. Dimastrodonato, L. O. Mereni, R. J. Young, and E. Pelucchi, "AlGaAs/GaAs/AlGaAs quantum wells as a sensitive tool for the MOVPE reactor environment," *J. Cryst. Growth*, vol. 312, no. 21, pp. 3057–3062, Oct. 2010.

# Experimental procedures

*PV device structure grown by MOVPE*

Table 2: Layer structure of as-grown GaAs photovoltaic laser power converters.

| Layer | Material | Alloy fraction | Thickness (nm) | Doping ($cm^{-3}$) | Role |
|---|---|---|---|---|---|
| 10 | GaAs | | 50 | p > $2 \times 10^{18}$ | Contact layer |
| 9 | $Al_xGa_{1-x}As$ | 0.1 | 150 | p > $1 \times 10^{18}$ | Current spreading |
| 8 | InGaP | Lattice matched | 30 | p - $1 \times 10^{18}$ | Front surface field |
| 7 | GaAs | | 150 | p - $1 \times 10^{18}$ | Emitter |
| 6 | GaAs | | 2000 | n - $2 \times 10^{17}$ | Base |
| 5 | $Al_xGa_{1-x}As$ | 0.3 | 50 | n - $1 \times 10^{18}$ | Back surface field |
| 4 | $Al_xGa_{1-x}As$ | 0.1 | 200 | n - $1 \times 10^{18}$ | Current spreading |
| 3 | GaAs | | 5 | n - $2 \times 10^{18}$ | Back contact |
| 2 | AlInP | Lattice matched | 500 | n - $1 \times 10^{18}$ | Release layer |
| 1 | $Al_xGa_{1-x}As$ | $0 \rightarrow 0.3$ | 50 | n - $1 \times 10^{18}$ | Buffer |
| 0 | GaAs | | substrate | n - x $2 \times 10^{18}$ | 2 deg to [110] |

*EQE measurements*

The EQE was measured in a Bentham PVE300 system. It uses two halogen light sources, Xenon and Quartz, through a monochromator to illuminate the device over a range of wavelengths. The light is optically chopped and the generated signal in the range of nanoamperes is recorded by a lock in amplifier.

*Current-voltage measurements*

The illuminated current-voltage setup used two probes, a source meter, a vacuum mount, a Keithley 2400 source measuring unit and high-powered red laser. An optical fiber is used as a waveguide to focus the light on to the active region of the PV devices. The probes and laser mount are connected to two separate Keithley SMUs. The laser was controlled with a second Keithley 2400 SMU to supply the current to the laser in order to control the output power of the laser. A fiber-coupled Sheaumann 808 nm multimode laser was used to illuminate the PV devices. The beam width from the fiber was 200 μm. To align the setup, the laser was approximately aligned over the center of the device with the help of a camera. A low current just above the threshold current of the laser was supplied to the laser. The position of the optical fiber was adjusted using x-y probe manipulators to maximize the measured photocurrent from the specific device which was displayed on the front panel of Keithley SMU. Once the photocurrent was maximized the fiber was assumed to be correctly aligned. The output power of the light exiting the fiber itself was measured using a Thorlabs

S415C thermopile photodetector. Once the optical fiber was aligned the SMU performed a voltage sweep while recording the photocurrent from the device. The voltage sweep was generally from -1V to 1.6V as the effect of the shunt resistance influenced the photocurrent up to -1V in the negative bias regime particularly for higher incident optical power densities. A voltage sweep was performed for varying incident power densities.